\documentclass{aa}
\usepackage{graphicx}
\usepackage{subfigure}
\usepackage{3424}

\begin{document}
\title{The line-of-sight distribution of water in the SgrB2 complex}

   \author{C.~Comito\inst{1}
          \and
          P.~Schilke\inst{1}
          \and
          M.~Gerin\inst{2}
          \and
          T.G.~Phillips\inst{3}
          \and
          J.~Zmuidzinas\inst{3}
          \and
          D.C.~Lis\inst{3}
          }
   \offprints{C. Comito\\ \email{ccomito@mpifr-bonn.mpg.de}}

   \institute{Max-Planck-Institut f\"ur Radioastronomie, Auf dem H\"ugel 69,
     D-53121 Bonn, Germany
     \and
     Laboratoire de Radioastronomie Millim\'etrique, Observatoire de Paris and 
     Ecole Normale Sup\'erieure, 
     24 rue Lhomond, F-75231 Paris, CEDEX 05, France
     \and
     California Institute of Technology, Downs Laboratory of Physics 320-47, 
     Pasadena, CA 91125, USA
     }

   \date{Received 16 December 2002/ Accepted 28 February 2003}

   \abstract{ We report the detection, with the Caltech Submillimeter
     Observatory\footnote{The CSO is operated by the California
       Institute of Technology under funding from the National Science
       Foundation, Grant No. AST-9980846.}, of the 894-GHz
     HDO($1_{1,1}-0_{0,0}$) transition, observed in absorption against
     the background continuum emission of the SgrB2 cores M and N.
     Radiative transfer modeling of this feature, together with the
     published data set of mm and submm HDO and \WATEIGH\ transitions,
     suggests that ground-state absorption features from deuterated
     and non-deuterated water trace different gas components along the
     line of sight. In particular, while the HDO line seems to be
     produced by the large column densities of gas located in the
     SgrB2 warm envelope, the \WATEIGH\ ground-state transition
     detected by SWAS and KAO at 548 GHz (Neufeld et al.
     \cite{neufeld2000}; Zmuidzinas et al. \cite{zmuidzi95a}) is
     instead a product of the hot, diffuse, thin gas layer lying in
     the foreground of the SgrB2 complex.
     
      }

   \maketitle
%

\section{Introduction}\label{intro}
The SgrB2 cloud is one of the most massive star forming regions in our
Galaxy.  It has many unique characteristics, among them an exceptional
chemistry.  Several species (FeO, Walmsley et al. \cite{walmsley2002};
NH$_2$, van Dishoeck et al. \cite{vandishoeck1993}; HF, Neufeld et al.
\cite{neufeld1997}; NH, Goicoechea, Cernicharo \& Caux
\cite{goicoechea2000}) have, in spite of searches elsewhere, been
detected only toward this source.  One possible explanation for this
enigmatic chemistry is the existence of a foreground layer of hot gas,
which has been observed e.g. in high-excitation ammonia transitions
(cf. H\"uttemeister et al.  \cite{huette1995}).  However, since this
layer has the same velocity as the ambient gas in SgrB2, it has been
proven difficult to assess its importance for the chemistry of many
species.  In particular its importance for the water chemistry has
been a matter of recent debate (Ceccarelli et al.
\cite{ceccarelli2002}, hereafter C02; Neufeld et al.,
\cite{neufeld2003}).  While its temperature and density are well
constrained by observational data (cf.  H\"uttemeister et al.
\cite{huette1995}), its column density and spatial extent have
remained elusive.  In this paper we are able to determine its
importance for water chemistry and provide constraints on its column
density and spatial width, by modeling the HDO and \WATEIGH\ emission
and absorption.


Water is known to be a fundamental ingredient of the interstellar
medium.  It is a major coolant of star-forming clouds (Ceccarelli,
Hollenbach \& Tielens, \cite{ceccarelli1996}) and therefore it affects
the dynamical evolution of the clouds. A ubiquitous tracer of
shock-heated gas, it dramatically influences the chemistry in shocked
regions (cf.  Neufeld \& Melnick \cite{neumel87}; Bergin, Melnick \&
Neufeld \cite{bergin1998}). Although direct, ground-based observations 
of non-masing
water lines are made extremely
difficult by atmospheric absorption, \WAT\ abundances can be estimated
via observations of isotopomers, such as \WATEIGH\ (e.g.  Phillips et
al. \cite{phillips1978}; Jacq et al.  \cite{jacq1988} and
\cite{jacq1990}; Gensheimer et al.  \cite{gensh1996}), and, in recent
years, by the availibility of satellites such as ISO (e.g.  van
Dishoeck \& Helmich \cite{vandihelm96}; Cernicharo et al.
\cite{cernicharo1997}; Wright et al. \cite{wright2000}) and notably
SWAS (cf.  Melnick et al.  \cite{melnick2000}; Snell et al.
\cite{snell2000a} and \cite{snell2000b}; Neufeld et al.
\cite{neufeld2000}, hereafter N00), but also by airborne observations
of H$_2^{18}$O (e.g.  Zmuidzinas et al. \cite{zmuidzi95a}, hereafter
Z95a; Timmermann et al.  \cite{timmermann1996}).  However, the
deuterated counterpart of water, HDO, presents many features
observable from ground in the cm, mm and submm wavelength atmospheric
windows (cf.  Henkel et al.  \cite{henkel1987}; Jacq et al.
\cite{jacq1990}; Schulz et al.  \cite{schulz1991}; Helmich, van
Dishoeck \& Jansen \cite{helmich1996}; Jacq et al.  \cite{jacq1999};
Pardo et al.  \cite{pardo2001}), and can also be used as a tracer of
water abundance under the assumption that both the deuterated and
non-deuterated species are spatially coexistent, and that their
abundance ratio is constant throughout the region of interest.

We report here the detection, with the Caltech Submillimeter Observatory,
of the ground-state ($1_{1,1}-0_{0,0}$) transition of deuterated
water, observed \emph{in absorption} against the background continuum
sources SgrB2(M) and SgrB2(N). The radial velocity of the observed HDO
feature suggests their direct connection to the SgrB2 complex.

In \S~\ref{subsect:coldens} we estimate the column density of the
absorbing HDO using classical absorption line assumptions, based on
the 894-GHz feature. Moreover, thanks to the \WATEIGH\ SWAS data of
N00, we are able to estimate the [HDO]/[\WAT] ratio towards the SgrB2
complex (\S~\ref{subsect:hdo_water_ratio}). The interpretation of
these results, however, depends critically on the assumptions on the
location of the absorbing gas. In fact, the observed absorption could
be produced both in the above-mentioned hot gas layer, and in the warm
envelope of molecular gas in which the two main cores are known to be
embedded.  Moreover, the assumption that deuterated and non-deuterated
water be spatially coexistent is probably incorrect in the SgrB2
cloud.  In \S~\ref{where}, we make use of our CSO 894-GHz data, as
well as of the published HDO and \WATEIGH\ observations carried out at
mm and submm wavelengths, to model the distribution of water
throughout the whole SgrB2 complex in its three component: hot cores,
warm envelope and hot layer, using a state-of-the-art radiative
transfer model. The results are discussed in \S~\ref{sec:discussion}.

\section{Observations}\label{section:obs}

The observations were carried out at the Caltech Submillimeter
Observatory atop Mauna Kea, Hawaii, on July 28, 2001. Scans were
taken, using the chopping secondary with a throw of 4$\arcmin$,
towards the coordinates $\alpha_{\rm{J2000}}=\rm{17^h47^m20\fs206}$,
$\delta_{\rm{J2000}}=\rm{-28\degr23\arcmin05\farcs27}$ for SgrB2(M),
and $\alpha_{\rm{J2000}}=\rm{17^h47^m20\fs389}$,
$\delta_{\rm{J2000}}=\rm{-28\degr22\arcmin22\farcs25}$ for SgrB2(N).
At the observing frequency of 893.6 GHz, the CSO 10.4-m antenna has a
HPBW of about 10$\arcsec$.

The 500-MHz facility AOS was used as backend, providing a velocity
resolution of $\sim 0.5$~\kms. System temperatures varied between 7500 and
8000 K. Several different LO settings were needed in order to cover a
wide enough velocity range, and to access at least two, namely the 65-
and 81-\kms, of the several velocity components known to exist in the
gas clouds lying on the line of sight in the direction of the two
cores (e.g., Whiteoak \& Gardner \cite{whiteoak1979};
Mart\'{\i}n-Pintado et al. \cite{mpintado1990}; Greaves et al.
\cite{greaves1992}; Tieftrunk et al. \cite{tieftrunk1994}). Also, the
use of various LO settings allowed us to rule out possible
contamination from signal coming from the image sideband. In total,
the range $-60 \, \kms \, \la v_{\rm LSR} \la 140 \, \kms\,$ was
covered.  However, the low signal-to-noise ratio does not allow an
analysis of the other velocity components, e.g. at $\sim0$ and
$\sim100$ \kms, towards which water absorption has been detected (cf.
N00).

We would also like to mention that the second ground-state rotational
transition of HDO, the 1$_{0,1}$-0$_{0,0}$ at 465 GHz, has been
searched for by us as well as by other authors (e.g., E.  Bergin and
collaborators, priv. comm.). A clear detection could not be achieved,
however all data sets are consistent in indicating that the 465-GHz
transition shows up as a weak \emph{emission} line.

Pointing is a delicate issue for the 850-GHz receiver (Kooi et al.
\cite{kooi2000}) at the CSO, therefore it was checked rather often
through 5-point maps of Mars, located only $\sim 11\degr$ away from
our target sources.  The accuracy of our pointing is confirmed by the
continuum levels we measure towards the two cores (21~K for M and 13~K
for N, in units of main-beam temperature), that match, within 20\%
error, with the values inferred by a map of the 350-$\rm{\mu m}$
continuum emission acquired with SHARC\footnote{\it{Submillimeter High
    Angular Resolution Camera}.} at the CSO (Dowell et al.
\cite{dowell1999}). Observing Mars also allowed us to measure the beam
efficiency of the telescope, which we found to be around 30\%.


\section{Results}

\subsection{Determination of HDO column densities}\label{subsect:coldens}

\begin{figure}[htbp]
\centering
\includegraphics[width=6cm,angle=-90]{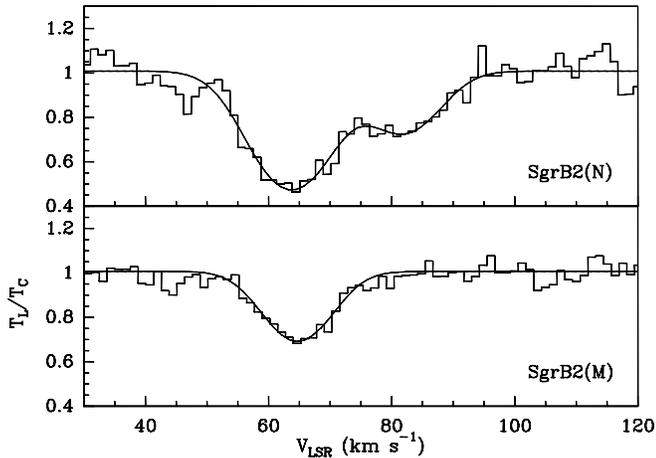}
\vspace{0.5cm}
\caption{The 894-GHz HDO ($1_{1,1}-0_{0,0}$) transition, observed in 
  absorption against the continuum emission of SgrB2(M) (lower panel)
  and of SgrB2(N) (upper panel). The $y$-scale gives the
  line-to-continuum ($T_{\rm L}/T_{\rm C}$) ratio. The absorption towards
  SgrB2(M) shows one velocity component only, around 65 \kms, while
  the gas in front of SgrB2(N) shows an additional component around 81
  \kms. Such behaviour has been found for other molecular species as
  well (e.g.  H$_2$CO, see Mart\'{\i}n-Pintado et al.
  \cite{mpintado1990}). The Gaussian fits to the optical depth of the
  lines are shown as solid curves, and the parameters derived from the
  fit are summarized in
  Tab.~\ref{tab:fit_results}.}\label{fig:sgrb2m_n}
\end{figure}

Fig.~\ref{fig:sgrb2m_n} displays two spectra of the ground-state
$1_{1,1}-0_{0,0}$ transition of HDO, observed in absorption against
the continuum emission of SgrB2(N) (upper panel) and SgrB2(M) (lower
panel).

The determination of column densities from absorption lines is in
general more accurate, with respect to the quantities inferred from
the analysis of emission lines, because the calculated value does not
depend on the excitation temperature of the line, $\rm{T_{ex}}$, as
long as this is negligible with respect to the temperature of the
background continuum source, $\rm{T_C}$: in other words, no
assumptions need be made about the physical state of the gas, except
that $\rm{T_{ex}} \ll T_C$ and that only the ground state of the
molecule is populated (we will show in \S~\ref{where} that neither
holds for a large part of the warm envelope). The total column density
of the absorbing species is given by:
\begin{equation}\label{eq:coldens}
N_{\rm tot} = \frac{8 \pi \nu^3}{A_{\rm ul} c^3} \Delta v \frac{g_{\rm l}}{g_{\rm u}} \tau.
\end{equation}
The opacity of an absorption line can be determined by:
\begin{equation}
\tau = -{\rm ln}\biggl[1-\frac{T_{\rm L}}{T_{\rm C}}\biggr],
\end{equation}
where $T_{\rm L}/T_{\rm C}$ is the line-to-continuum ratio. Therefore,
eq.~(\ref{eq:coldens}) only depends on $T_{\rm L}/T_{\rm C}$ and on
the line width, $\Delta v$, quantities that can be easily derived
from a Gaussian fit of the optical depth. Note that, at this stage, no
assumptions have been made about the actual location (warm envelope or
hot layer) of the gas producing the observed features.

The fit to the 65- and 81-\kms\ components of N, and to the 65-\kms\ 
component of M are shown in Fig.~\ref{fig:sgrb2m_n}, while the derived
parameters are summarized in Tab.~\ref{tab:fit_results}. The relative
values of the total column density are listed in
Tab.~\ref{tab:m_n_coldens}. Studies of the 1.3-mm dust emission in the
region (IRAM 30-m maps, Gordon et al. \cite{gordon1993}) indicate that
the column density of H$_2$, $N({\rm H_2})$, reaches values of $2.2
\times 10^{24}$ \cmsq\ towards SgrB2(M), and $7.5 \times 10^{24}$
\cmsq\ towards SgrB2(N). Such results agree, within a factor of 2,
with previous estimates obtained by Goldsmith, Snell \& Lis
(\cite{goldsmith1987}) based on lower-spatial-resolution 1.3-mm
continuum maps. The relative abundance of gas-phase HDO displayed in
the last column of Tab.~\ref{tab:m_n_coldens} has been estimated on
the basis of the above mentioned values of $N({\rm H_2})$. The
estimate takes into account that the amount of gas we ``see'' through
the HDO absorption is only a half of the actual amount of gas measured
through the core. We take the error on the HDO abundance to be of
$\sim 50 \%$.

\begin{table}[htbp]
\caption{Results of the fits to the optical depth of the HDO absorption features shown in
Fig.~\ref{fig:sgrb2m_n}. Columns 2 to 4 give, respectively, the radial 
velocity of the absorbing gas, the line opacity 
and the line width. The fits have been performed with the method ABSORPTION from the GILDAS CLASS package.
}
\label{tab:fit_results}
\begin{center}
\begin{tabular}{cccc}
\hline
\hline
            SOURCE  & $\rm{v_{LSR}}$ & $\tau$ & $\rm{\Delta v}$ \\
                    & (\kms)         &                  & (\kms) \\
            \hline
            SgrB2(M)& $64.7 \pm 0.5$        & $0.38 \pm 0.04$  & $12.5 \pm 1.3$  \\
            SgrB2(N)& $63.7 \pm 0.5$        & $0.77 \pm 0.05$  & $14.6 \pm 1.3$  \\
            SgrB2(N)& $82.1 \pm 0.9$        & $0.32 \pm 0.04$  & $13.0 \pm 1.7$  \\
\hline
\end{tabular}
\end{center}
\end{table}

\begin{table}[htbp]
  \caption{Column densities of HDO calculated from the absorption features shown in 
Fig.~\ref{fig:sgrb2m_n}. Column 2 gives the radial velocity of the 
absorbing gas component. The values of the column density (column 3) are derived, 
using eq.~(\ref{eq:coldens}), assuming that only the ground-state level of HDO is populated.
Column 4 lists the estimated HDO abundance, calculated assuming a H$_2$ column density of
$2.2 \times 10^{24}$ \cmsq\ towards SgrB2(M) and $7.5 \times 10^{24}$ \cmsq\ towards SgrB2(N)
(Gordon et al. \cite{gordon1993}). The estimated error of the HDO abundance is of order 50\%.}
  \label{tab:m_n_coldens}
  \begin{center}
    \begin{tabular}{cccc}
\hline
\hline
            SOURCE  & $\rm{v_{LSR}}$ & $\rm{N(HDO)}$    &  [HDO] \\
                    &     (\kms)     & ($\times 10^{13}$ \cmsq) &   ($\times 10^{-11}$) \\
\hline
            SgrB2(M)& $64.7 $        & $1.2 \pm 0.2$            & $ 1.1$    \\
            SgrB2(N)& $63.7 $        & $2.8 \pm 0.3$            & $ 0.7$    \\
            SgrB2(N)& $82.1 $        & $1.1 \pm 0.2$            & $ 0.3$    \\
\hline
    \end{tabular}
  \end{center}
\end{table}

\subsection{The [HDO]/[H$_2$O] ratio}\label{subsect:hdo_water_ratio}

N00 have observed, with the {\it Submillimeter Wave Astronomy
  Satellite} (SWAS), the ground-state $1_{1,0}-1_{0,1}$ transition of
o-\WATEIGH\ (547.7 GHz), in absorption against the continuum emission
of SgrB2. They were therefore able to estimate the column density of
\WATEIGH\ over a wide range of radial velocities (from -120 to 20
\kms).

We use the \WATEIGH\ spectrum of N00 to estimate the [HDO]/[\WAT]
ratio in the gas components with radial velocities of 63 and 81 \kms.
The comparison between the two data sets is not straightforward, since
the spatial resolution of SWAS, $3\farcm3 \times 4\farcm5$, is such
that the emission from SgrB2(M) and SgrB2(N), which are $\sim
47\arcsec$ apart, is not resolved. To get around this problem, we
simply sum up the HDO scans towards M and N weighted by the
attenuation of the circularized $3\farcm9$ SWAS beam, and compare the
resulting spectrum (hereafter M+N) to that of N00.  Such procedure is
based on the assumption that the absorbing cloud is extended and
uniform over the SWAS beam size. This is a reasonable approximation
for the 65-\kms\ component, which is known to be extended over both
SgrB2(M) and N.  However, the 81-\kms\ gas component is definitely
smaller than the SWAS beam (it is only observed towards the northern
core).

\begin{figure}[htbp]
  \centering 
  \includegraphics[width=6.cm,angle=-90]{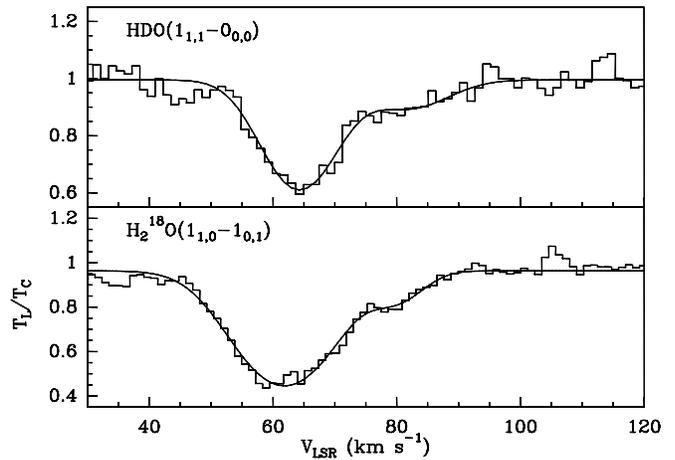}
  \vspace{.5cm}
\caption{The 894-GHz HDO($1_{1,1}-0_{0,0}$) and 
  the 548-GHz o-\WATEIGH($1_{1,0}-1_{0,1}$) absorption features
  observed towards SgrB2. The spectrum in the {\it upper panel} is the
  sum of the scans relative to the HDO absorption towards SgrB2(M) and
  SgrB2(N) separately (see Fig.~\ref{fig:sgrb2m_n}), observed at the
  CSO with a $10\arcsec$ beam. The {\it lower panel} shows the
  o-\WATEIGH\ absorption towards SgrB2, observed by N00 with SWAS
  ($3\farcm3 \times 4\farcm5$ beam). The $y$-scale gives the
  line-to-continuum ($\rm{T_L/T_C}$) ratio. The solid curves represent
  the Gaussian fits to the line opacity. The radial velocities, line
  opacities and widths, derived from the fit for both transitions, are
  listed in Tab.~\ref{tab:m+n_fit}. The larger width of the SWAS
  o-\WATEIGH\ line, with respect to the HDO feature resulting from the
  combination of the CSO scans, agrees with the extended water
  distribution observed by, e.g., Cernicharo et al.
  (\cite{cernicharo1997}) and Neufeld et al.
  (\cite{neufeld2003}).}\label{fig:hdo_water}
\end{figure}

Fig.~\ref{fig:hdo_water} compares the HDO absorption features towards
M+N (upper panel) with the 548-GHz o-\WATEIGH\ ground-state line
observed by N00 (lower panel). The solid curves in
Fig.~\ref{fig:hdo_water} represent the two-component Gaussian fits to
the opacity of the observed features, and the fitting parameters are
summarized in Tab.~\ref{tab:m+n_fit}. Again, we used
eq.~(\ref{eq:coldens}) to calculate the total column density of the
two species, assuming that, for both of them, only the ground-state
level is populated. The results are summarized in
Tab.~\ref{tab:m+n_coldens}. The last column of
Tab.~\ref{tab:m+n_coldens} shows the estimated value of the
[HDO]/[\WAT] ratio for the two velocity components of the gas,
calculated assuming that $[\SIXO]/[\EIGHO] = 261\pm20$ (Whiteoak \&
Gardner \cite{whiteoak1981}), and that [o-\WAT]/[p-\WAT]$=3$.  We find
the [HDO]/[\WAT] ratio towards SgrB2, at 63 and 81 \kms, to be $\sim
5\times10^{-4}$ and $\sim 10^{-3}$ respectively (see
Tab.~\ref{tab:m+n_coldens}). These values are about 30 and 70 times
higher than the measured deuterium abundance in the Local Interstellar
Medium (1.5$\times10^{-5}$, Linsky \cite{linsky1998}). The enhancement
of the deuterium fractionation observed in the [HDO]/[\WAT] ratio
appears even more important if one considers that deuterium has been
measured to be under-abundant in the Galactic Center region.  Lubowich
et al.  (\cite{lubowich2000}) have estimated the [D]/[H] ratio towards
SgrA to be around 1.7$\times10^{-6}$. Similar values have been
measured towards SgrB2 by Jacq et al. (\cite{jacq1999}) and by
Polehampton et al. (\cite{polehampton2002}). All estimates are
affected by very large uncertainties, however an order-of-magnitude
comparison with our measured [HDO]/[\WAT] ratio shows that it is
actually a few $10^2$ times higher than the deuterium abundance in the
region. This value, scaled to the lower [D]/[H] ratio in the Galactic
Center, is consistent with that predicted by the steady-state chemical
models of Roberts \& Millar (\cite{robmill2000}) for gas temperatures
ranging between $\sim30$ and 100~K and densities between $\sim 10^3$
and $10^8$ \percc, roughly the range of temperatures and densities
expected in the warm envelope. However, the analysis presented in the
next section shows that the HDO and \WAT\ absorption are actually
produced in different locations, hence the above mentioned value of
the [HDO]/[\WAT] ratio has little relevance. We will address the issue
again in \S~\ref{sec:discussion}.

\begin{table*}[htbp]
\caption{Results of the Gaussian fits to the optical depth of the HDO (columns 1-3) and o-\WATEIGH\ 
  (columns 4-6, N00 data) absorption features observed towards SgrB2 
  and shown in Fig.~\ref{fig:hdo_water}. The listed parameters are, for each transition, 
  respectively the radial velocity, the line opacity and the line width. The fits have been performed with 
the method ABSORPTION from the GILDAS CLASS package. }

\label{tab:m+n_fit}
  \begin{center}
    \begin{tabular}{ccccccc}
            \hline
            \hline
              \multicolumn{3}{c}{HDO(1$_{1,1}-0_{0,0}$)}&  & \multicolumn{3}{c}{o-\WATEIGH($1_{1,0}-1_{0,1}$)} \\
        \cline{1-3} \cline{5-7}
   $\rm{v_{LSR}}$ & $\tau$         & $\rm{\Delta v}$ & &  $\rm{v_{LSR}}$ &  $\tau          $ & $\rm{\Delta v}$ \\
    (\kms)      &                  &  (\kms)         & &      (\kms)     &                   &  (\kms)         \\
\cline{1-3} \cline{5-7}
 $64.3\pm0.3$   &  $0.49\pm0.03$   & $13.3\pm0.7$    & &  $61.9\pm0.3$   &    $0.77\pm0.02$  &  $17.7\pm0.7$   \\
 $82.0\pm0.0$   &  $0.11\pm0.02$   & $14.6\pm0.2$     & &  $80.3\pm0.9$   &    $0.13\pm0.02$  &  $9.6\pm2.7$    \\
            \hline
    \end{tabular}
  \end{center}
\end{table*}

\begin{table*}
\caption{Total column densities of HDO and o-\WATEIGH\ in the $3\farcm3 \times 4\farcm5$ SWAS beam, 
producing the absorption features shown in Fig.~\ref{fig:hdo_water}. Column 1: average radial velocity of the
gas component (calculated from the values in Tab.~\ref{tab:m+n_fit}, columns 1 and 4); column 2 and 3:
total column density of HDO and o-\WATEIGH\, calculated using eq.~\ref{eq:coldens} and the fit parameters 
in Tab.~\ref{tab:m+n_fit}. Since we assumed only the ground-state level the absorbing material 
to be populated, these are to be considered as lower limits. 
Column 4: [HDO]/[\WAT] ratio, calculated on the basis of the estimated 
column densities of HDO and o-\WATEIGH\ assuming that $[\SIXO]/[\EIGHO] = 261\pm20$ 
(Whiteoak \& Gardner \cite{whiteoak1981}) and that [o-\WAT]/[p-\WAT]$=3$.
}

\label{tab:m+n_coldens}
  \begin{center}
    \begin{tabular}{ccccc}
      \hline
      \hline
      $\rm{v_{LSR}}$   &  N$({\rm{HDO}})$ & $\rm{N(o-\WATEIGH)}$ &  $\rm{N(\WAT)}$ &  [HDO]/[\WAT]\\
         (\kms)        &   ($10^{13}$ \cmsq)&        ($10^{13}$ \cmsq)  &    ($10^{16}$ \cmsq)  & ($10^{-3}$)      \\
         \hline
         $\sim 63$     &  $1.67\pm0.13$       &   $  6.33\pm0.30$     &  $2.20\pm0.20$       &   $0.8\pm0.1$      \\
         $\sim 81$     &  $0.40\pm0.07$       &   $  0.60\pm0.19$     &  $0.21\pm0.07$       &   $1.9\pm0.7$      \\
\hline
    \end{tabular}
  \end{center}
\end{table*}

\section{Modeling the distribution of water}\label{where}

As previously mentioned (\S~\ref{intro}), it is known from
observations of, e.g., H$_2$CO (cf. Whiteoak \& Gardner
\cite{whiteoak1979}; Mart\'{\i}n-Pintado et al.  \cite{mpintado1990})
that the two main cores in the SgrB2 complex - SgrB2(N) and SgrB2(M) -
are embedded in a warm (15~K~$\la T_{\rm gas}\la 100$~K), dense
($n({\rm H_2})\sim 10^5$ \, \percc) envelope of molecular gas. Also, a
hot (500~K~$\la T_{\rm gas}\la 700$K) and diffuse ($10^3 \, \percc \la
n({\rm H_2})\la 10^4 \, \percc$) gas layer is located in the
foreground, as indicated by the detection in absorption of high-energy
ammonia transitions (Wilson et al.  \cite{wilson1982}; H\"uttemeister
et al.  \cite{huette1995}; C02).  Both gas components show
characteristic radial velocities around 65 and 81 \kms (the latter
being observable only towards the northern core), hence the hot
foreground gas layer is thought to be physically connected to the
cloud complex as well.

The interpretation of the results illustrated in
$\S~\ref{subsect:hdo_water_ratio}$ strongly depends on the assumptions
about the location of the absorbing gas. Absorption due to molecules
such as \WAT\ and its isotopomers has generally been attributed to the
warm envelope (Z95a; N00), under the assumption that only the
ground-state level was populated.  However, C02 have proposed that all
the absorbing water might instead be located in the hot gas layer. In
this case, according to C02, the estimated column density of \WAT\ 
towards the main velocity components would be about one order of
magnitude higher than that quoted, for example, by Z95a and by us in
\S~\ref{subsect:hdo_water_ratio}. Such a high value would support the
C-shock model used by Flower and collaborators to explain the observed
features of the hot layer, such as the high gas temperature, low gas
density and limited spatial width (Flower \& Pineau des For\^ets
\cite{flowpin1994}; Flower, Pineau des For\^ets \& Walmsley
\cite{flower1995}).

From a chemical point of view, it is entirely possible that a large
percentage of the observed water column density be located within the
diffuse hot layer rather than in the warm envelope. In fact, the gas
temperature in the hot layer is sufficient not only to induce
evaporation of water ices from the surface of dust grains, but also to
trigger the gas-phase production of \WAT\ via the following chain of
neutral-neutral reactions:
\begin{eqnarray*}
{\rm O + H_2} & \rightarrow & {\rm OH + H}, \\
{\rm OH + H_2} & \rightarrow & {\rm H_2O + H},
\end{eqnarray*}
which proceed very rapidly as $T \ga 400$~K (Elitzur \& de Jong
\cite{elitzur1978}, also cf. Bergin, Melnick \& Neufeld
\cite{bergin1998}). Therefore, we can expect the abundance of \WAT\ to
be enhanced in the hot layer with respect to the warm envelope, up to
the point where all the oxygen not locked in CO is converted into
\WAT.  In fact, HDO can be formed in shocked gas as well, via the
reactions (Bergin, Neufeld \& Melnick (\cite{bergin1999}):
\begin{eqnarray*}
{\rm O + HD} & \rightarrow & {\rm OD + H}, \\
{\rm OD + H_2} & \rightarrow & {\rm HDO + H},
\end{eqnarray*}
and
\begin{eqnarray*}
{\rm OH + HD} & \rightarrow & {\rm HDO + H}.
\end{eqnarray*}
But, because of the lower HD abundance, the timescales are slower for
these reactions than for the \WAT\ production paths, such that all the
available atomic oxygen will be locked in \WAT\ prior to the formation
of HDO. Thus, gas-phase chemistry in the hot layer yields
[HDO]/[\WAT]~$\sim$~[D]/[H]. We believe this picture cannot be
significantly altered by grain chemistry, because {\it i)} the
temperature of the dust in the hot layer is known to be much lower
than that of the gas (cf. Wilson et al.  \cite{wilson1982},
H\"uttemeister et al. \cite{huette1993}), too low to allow evaporation
of ice mantles from the grain surface; {\it ii)} sputtering of dust
grains, on the other hand, cannot account for the release of intact
molecules; and however, {\it iii)}, the build-up of ice mantles is not
efficient at low densities, hence the deposit of water ices on the
grains surface is likely to be negligible in the first
place\footnote{This is supported, for example, by observations of high
  abundances of gas-phase atomic oxygen in molecular clouds with
  density $\sim 10^2 - 10^3$~\percc\ (Lis et al.  \cite{lis2001}),
  suggesting that much of the oxygen which is not locked in CO could
  well be present in the gas in its atomic form, rather than depleted
  on dust grains.}.

To summarize, we believe that it is reasonable to expect the
contribution of the hot layer to the observed HDO absorption to be
negligible, whereas its contribution to the observed \WAT\ absorption
is likely to be significant. It is important to estimate the extent of
such a contribution: a precise determination of the column density of
water would set a tight constraint on the physical models that attempt
to identify the heating mechanism responsible for the high
temperatures in the hot layer. Unfortunately, since both gas
components show roughly the same radial velocity, it is not possible
to separate, by purely observational means, the ground-state water
absorption produced in the warm envelope from that produced in the hot
layer. However, a number of HDO and \WATEIGH\ transitions have been
observed towards the SgrB2 cores with a variety of instruments (see
Tab.~\ref{tab:data_list} and Fig.~\ref{hdo_levels}), and detailed
modeling can be performed to disentangle the contribution to the
observed features from the different cloud components.

The available dataset, enriched by our detection of the ground-state
HDO transition at 894 GHz, provides sufficient observational
constraints to model, in a self-consistent manner, the HDO and \WAT\ 
abundance in all three components of the SgrB2 cloud. In detail:

\begin{figure}
\centering
\includegraphics[angle=0,width=9cm]{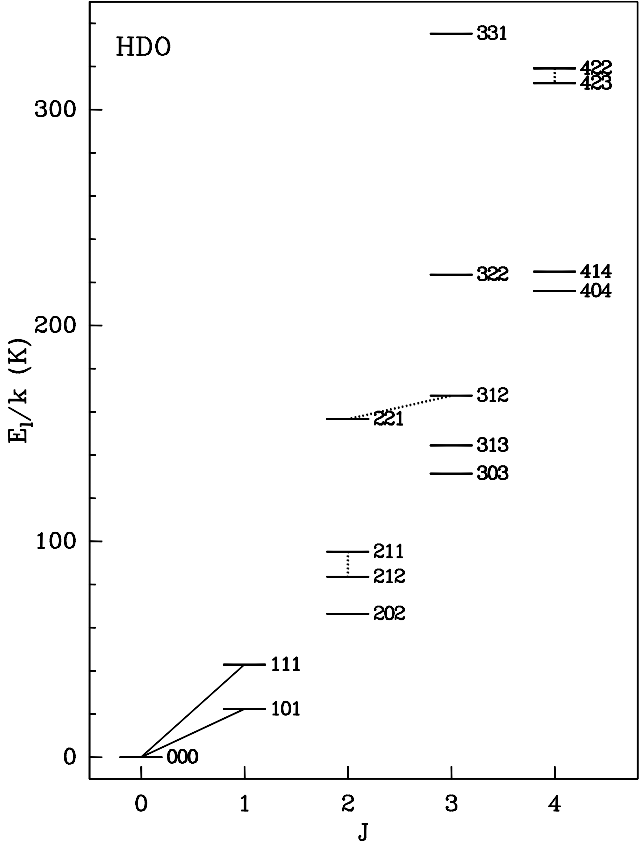}
\caption{Rotational levels of deuterated water up to $\sim 400$~K. 
  The transitions included in our data set are indicated (solid lines:
  this work; dotted lines: Jacq et al. \cite{jacq1990}; see
  Tab.~\ref{tab:data_list}).  }\label{hdo_levels}
\end{figure}

\begin{itemize}
\item[{\it a)}] the observed \emph{emission} lines of p-\WATEIGH\ at
  203 GHz (SgrB2(N), Gensheimer et al.  \cite{gensh1996}), and of HDO
  at 143, 226, 241 GHz (SgrB2(M) and N, Jacq et al. \cite{jacq1990})
  constrain the HDO and water abundance and the H$_2$ density in the
  hot-core-type components;
\item[{\it b)}] being virtually absent in the hot layer, HDO can be
  used to calibrate the water content of the warm envelope. In
  particular, the 894-GHz HDO \emph{absorption} feature (this work)
  observed towards both SgrB2(M) and N, together with the [HDO]/[\WAT]
  ratio estimated at point {\it a)}, help us set the HDO and \WAT\ 
  abundance in this region;
\item [{\it c)}] finally, having estimated the column density of water
  in the warm envelope, it is possible to assess whether or not the
  quantity of non-deuterated water in the warm envelope can be
  responsible for the \WATEIGH\ absorption observed at 548~GHz (N00,
  Z95a). The 180-$\mu$m o-\WATEIGH\ line observed by Cernicharo et al.
  (\cite{cernicharo1997}) towards the SgrB2 complex is blended with
  the H$_3$O$^+$(1$_1^-$-1$_1^+$) line (Goicoechea \& Cernicharo
  \cite{goicoechea2001}), and cannot therefore be used as a reference
  for our purposes.
\end{itemize}

\begin{table*}[htbp]
  \begin{center}
    \caption{Data set used to model the HDO and \WAT\ distribution in
      the SgrB2 cloud complex. References are: Jacq et al.
      \cite{jacq1990} (J90), Neufeld et al. \cite{neufeld2000} (N00),
      Gensheimer et al.  \cite{gensh1996} (G96). The set of HDO
      transitions is displayed in Fig.~\ref{hdo_levels}.
      }\label{tab:data_list}
    \begin{tabular}{lrcccl}
\hline
\hline
Transition                      & $E_{\rm l}/k$ & Frequency & HPBW & Observed   & Ref. \\
                                &  (K)          &  (GHz)    &      & towards    &  \\
\hline                                                                                
HDO(1$_{1,1}$-0$_{0,0}$)$^{\mathrm{a}}$&   0.0         & 893.6     &$11\arcsec$  & M, N       & this work (CSO) \\  
HDO(2$_{1,1}$-2$_{1,2}$)$^{\mathrm{b}}$& 83.6          & 241.6     &$10\arcsec$  & N          & J90 (IRAM 30m)       \\
HDO(3$_{1,2}$-2$_{2,1}$)$^{\mathrm{b}}$& 156.7         & 225.9     &$11\arcsec$  & M, N       & J90 (IRAM 30m)      \\
HDO(4$_{2,2}$-4$_{2,3}$)$^{\mathrm{b}}$& 312.3         & 143.7     &$17\arcsec$  & N          & J90 (IRAM 30m)      \\
o-\WATEIGH(1$_{1,0}$-1$_{0,1}$)$^{\mathrm{a}}$& 34.2   & 547.7     &$3\farcm9$   & M+N        & N00 (SWAS)      \\
p-\WATEIGH(3$_{1,3}$-2$_{2,0}$)$^{\mathrm{b}}$& 193.9  & 203.4     &$12\arcsec$  & N          & G96 (IRAM 30m)      \\
\hline
    \end{tabular}
\begin{list}{}{}
\item[$^{\mathrm{a}}$] absorption 
\item[$^{\mathrm{b}}$] emission
\end{list}
  \end{center}
\end{table*}  

\subsection{The model}\label{molnd}

We use the static radiative transfer code described by
Zmuidzinas et al.  \cite{zmuidzi95b} (hereafter Z95b) to reproduce the
intensities observed for the features listed in
Tab.~\ref{tab:data_list}. A few changes have been made with respect to
the original version, but the bulk of the model is the same and can be
summarized as follows:

\begin{itemize}
\item A background, internally heated continuum source is embedded in
  a spherical molecular envelope of radius $0.05$~pc$ \leq r \leq
  22.5$~pc (corresponding to the warm envelope as defined at the
  beginning of \S~\ref{where}).
\item The large-scale structure of the cloud is well constrained by
  the observations of C$^{18}$O carried out by Lis \& Goldsmith
  (\cite{lisgold1989}, hereafter LG89). The density profile adopted in
  the model, as well as the radial variation of the dust temperature,
  correspond to their Model C. The temperature of the gas is coupled
  to that of the dust, $T_{\rm dust} = T_{\rm gas} = T$.
\item The abundance of the modeled species, assumed to be constant
  throughout the envelope, is a free parameter, as well as the width
  (in \kms) of the modeled feature.
\item The code allows radiative transfer calculations for one Gaussian
  velocity component only. In what follows, we will only try to
  reproduce the 65-\kms\ component of the molecular gas.
\end{itemize}

This geometrically straightforward model has proved successful in
reproducing the dust continuum emission observed towards the SgrB2
cloud complex at mm and submm wavelengths, as well as absorption
features such as the HCl(1-0) and \WATEIGH($1_{1,0}-1_{0,1}$) lines
observed in the submm band with the Kuiper Airborne Observatory (KAO,
see Z95a and Z95b).  Nevertheless, it is not capable, as it is, to
reproduce those HDO and \WATEIGH\ features that have been observed in
emission, and that are thought to be produced in the hot-core-type
components of SgrB2(M) and SgrB2(N) (cf. Jacq et al. \cite{jacq1990},
Gensheimer et al.  \cite{gensh1996}). In fact, a \emph{molecular}
hot-core component, intended as a compact, dense, warm molecular
region characterized by fairly flat radial profiles of density and
temperature, is missing in the model, since the region of radius $r <
0.05$~pc is devoid of molecular gas. A radius of $\sim 0.05$~pc has
indeed been attributed to the hot cores embedded in SgrB2(M) and N,
based on the high-spatial-resolution interferometric maps of the
continuum emission towards the complex (cf. Lis et al.
\cite{lis1993}, hereafter L93).  Therefore, we have modified the model
by introducing a region of radius $r < 0.05$~pc, in which the H$_2$
density and the gas and dust temperatures are constant and equal the
values of density and temperature at $r = 0.05$~pc, based on the
radial profiles of LG89, Model C. This yields $n({\rm H_2}) \sim
3.4\times10^7$~\percc\ and $T_{\rm dust} = T_{\rm gas} \sim 200$~K in
the hot-core component.  Both values are in good agreement with the
estimates based on the continuum emission ($n({\rm H_2}) \sim 3\times
10^7$ \percc\ in the hot core of SgrB2(M), and $n({\rm H_2}) \sim
2\times 10^7$ \percc\ in SgrB2(N), L93) and on the intensities of
metastable and non-metastable ammonia lines ($T_{\rm kin} \sim T_{\rm
  rot} = 202 \pm 15$~K in SgrB2(M), Vogel, Genzel \& Palmer
\cite{vogel1987}). Note that the introduction of the tiny hot-core
component does not produce significant variations in the predicted
continuum emission.

\begin{figure}[htbp]
\includegraphics[angle=-90,bb=25 120 575 780,clip,width=9.5cm]{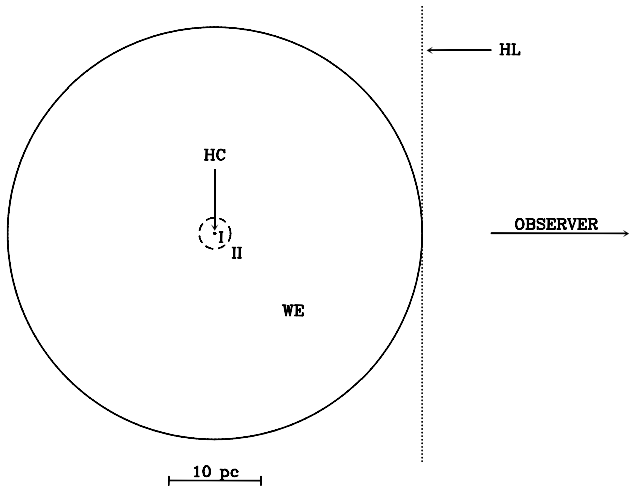}
\caption{Sketch illustrating the relative sizes of the SgrB2 cloud components. For 
  practical reasons, only one of the two cores is represented here.
  The hot core (labelled HC in the figure) has a radius of 0.05~pc,
  and is tiny compared to the size of the warm envelope (WE, $r =
  22.5$~pc).  The dashed circle identifies the transition radius ($r =
  1.7$~pc) between Phase I ($T > 100$~K) and Phase II ($T<100$~K).
  Modeling of the spherically symmetric regions is illustrated in
  detail in \S~\ref{molnd}. The hot layer (HL) is shown as a thin
  (0.02~pc wide) sheet of gas lying right outside the warm envelope
  (see \S~\ref{hotlayer}).  }\label{source_model}
\end{figure}

We also consider the effect of temperature on the gas-phase abundance
of the modeled species. A vast ($\sim 1.7$~pc~$ \leq r \leq 22.5$~pc)
portion of the warm envelope shows temperatures lower than 100~K,
which favour the freeze-out of gas-phase water onto dust grains (cf.
Williams \cite{williams1993} and references therein). The abundance of
water (both deuterated and non-deuterated) in this region will
naturally be lower than it would be at temperatures higher than 100~K,
and particularly in the vicinity of the hot cores, where the higher
$T$ promotes the \emph{evaporation} of water ices from the grains
surface. The second major modification to the Z95b model involves
therefore a differentiation between a warmer ($T > 100$~K, hereafter
Phase I\footnote{Note that Phase I includes, but is not limited to,
  the hot-core component of the cloud (see sketch in
  Fig.~\ref{source_model}).}) and a colder ($T < 100$~K, hereafter
Phase II) region, identified by different gas-phase abundances.  This
is of course an oversimplification, however such an approximation is
sufficient to reproduce the data correctly, within the errors, as will
be illustrated in \S~\ref{hotcore}.  As already discussed by Neufeld
et al. (\cite{neufeld1997}), who introduced a similar differentiation
in order to reproduce the \WAT($4_{3,2}-4_{2,3}$) emission line
observed with ISO at 122~$\mu$m, this freeze-out assumption is
required by the observations: a reduction of the HDO and \WAT\ 
abundance in the outer regions of the envelope is \emph{necessary} to
model the HDO and \WATEIGH\ emission radiated from the hot-core gas
and, at the same time, the HDO absorption thought to be produced in
the warm envelope. In fact, a decrease in the water abundance towards
the outer regions several molecular cloud cores has been observed by
Snell et al. (\cite{snell2000b}). The sketch in
Fig.~\ref{source_model} shows the relative sizes of hot core and warm
envelope, as well as of the Phase I and Phase II regions.

Finally, as indicated in Tab.~\ref{tab:data_list}, all the features in
our sample but the two o-\WATEIGH\ lines have been observed with
spatial resolutions ranging between $10\arcsec$ and $17\arcsec$.
Since the projected distance between the northern and the middle core
is of $\sim 47\arcsec$ ($\sim 2$~pc at a distance of 8.5 kpc), it is
more appropriate to model the two cores separately. In fact, M and N
show different chemical compositions, with the northern core
displaying higher abundances of complex molecules (cf. Snyder, Kuan \&
Miao \cite{snyder1994}; Miao \& Snyder \cite{miao1997} and references
therein; Nummelin et al.~\cite{nummelin2000}). The middle core, on the
other hand, presents a higher mass and luminosity, as inferred from
the analysis of the dust emission at several wavelengths (cf.
Goldsmith et al.  \cite{goldsmith1992}; L93; Gordon et al.
\cite{gordon1993}; Vogel, Genzel \& Palmer \cite{vogel1987}; Dowell et
al. \cite{dowell1999}).  Such dissimilarities seem to reflect the
youth of N relative to M. However, for our purposes we will neglect
the chemical differences between the two cores, since the absence of
molecular species in the middle core mostly concerns large complex
molecules (see references above). Moreover, our simple model shows
that the different continuum temperatures observed towards the two
cores can be reproduced fairly well by simply assuming different dust
properties for the two cores.  L93 suggested, for the grain emissivity
law in the northern core, a shallower slope than that of the middle
core, a difference that may be explained by different grain shapes in
the two cores.

We proceed by running our radiative transfer model separately for the
two sources. The results will be appropriately combined, when
necessary, to be compared to the data (see \S~\ref{hotcore}). We
assume identical physical parameters (density and temperature
profiles, size of the hot core and of the warm envelope, \WAT\ and HDO
abundance) for both M and N. The only difference between the two cores
lies in the grain emissivity law, that has a slope of 1.1 in the
northern core, and of 1.4 in the middle one (L93). This condition
holds only in the inner $\sim 1$~pc of the cloud, which roughly
correspond to half the projected distance between the two cores. In
fact, we must take into account the fact that the warm envelopes in
which the cores are embedded will, at some point, merge to form one
single ''shared'' envelope. Hence, for radii larger than half the
distance between the cores, we assume the dust properties to be the
same, with a grain emissivity slope of 1.5 as derived from submm flux
densities in a $60\arcsec$ beam (Goldsmith et al.  \cite{gold1990}).

\subsection{{\rm HDO} and \WAT\ in the hot cores and in the warm envelope}\label{hotcore}

\begin{figure}[htbp]
\includegraphics[angle=-90,width=8cm]{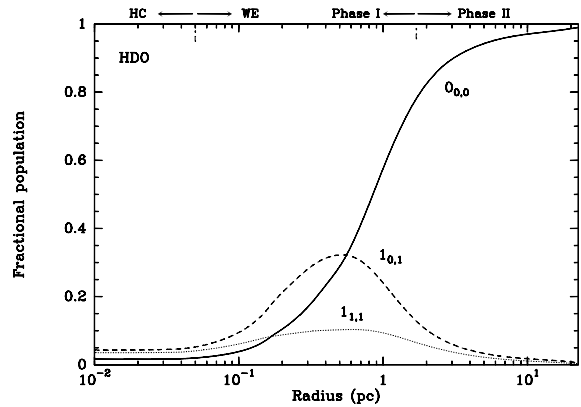}
\caption{Fractional populations of the three lowest-energy 
  levels of the HDO as a function of the radius of the cloud.
  }\label{hdo_pop}
\end{figure}

We aim at reproducing the 143-, 226-, and 241-GHz HDO emission
features, observed by Jacq et al. (\cite{jacq1990}, also see
Tab.~\ref{tab:data_list}), in order to model the gas-phase abundance
of this species in the hot-core components of SgrB2(M) and N. At the
same time, we rely on the 894-GHz ground-state absorption feature
(Fig.~\ref{fig:sgrb2m_n}) to constrain the HDO abundance in the warm
envelope. The closest match between model and data is obtained when
the abundance of HDO, relative to H$_2$, is set to $1.5\times10^{-9}$
in Phase I, and $3.5\times10^{-11}$ in Phase II of the molecular
cloud, i.e., a depletion of a factor of 40 is observed in Phase~II.
We note that this model predicts the 465-GHz line to be in weak
emission, in agreement with the observations (cf.
\S~\ref{section:obs}). The total column density of HDO in a
10$\arcsec$ beam, $N({\rm HDO})$, is as high as $2 \times
10^{16}$~\cmsq, almost three orders of magnitude higher than estimated
from the 894-GHz absorption line (\S~\ref{subsect:coldens}) under the
assumption that all the absorbing HDO is in the ground state. This
inconsistency is explained by the radial distribution of the
fractional populations of the first three levels of HDO ($0_{0,0}$,
$1_{0,1}$ and $1_{1,1}$ in order of increasing energy,
Fig.~\ref{hdo_pop}): the fractional population of the ground level
exceeds 90\% for $r \ga 3$ pc only, hence our calculations in
\S~\ref{subsect:coldens} severely underestimate the total HDO column
density in the innermost portion of the cloud, which shows the highest
gas density and water abundance and thus contributes the largest
percentage ($\geq 99$\%) of the total HDO column density.
Incidentally, note that our value of $N({\rm HDO})$ is only about one
order of magnitude higher than that estimated by Jacq et al.
(\cite{jacq1990}) towards SgrB2(N) through the analysis of the
hot-core transitions, that instead trace the densest gas in the
complex.

The abundance of non-deuterated water in the hot core can be modeled
on the basis of the 203-GHz p-\WATEIGH\ transition observed, in
emission, towards SgrB2(N) (Gensheimer et al. \cite{gensh1996}). We
are able to reproduce the measured intensity of this line if we assume
the abundance of p-\WATEIGH\ to be $2.2\times10^{-9}$ in Phase I.
Note that this feature shows, towards SgrB2(N), severe blending with a
wide SO$_2$ line, therefore the intensity indicated by Gensheimer et
al. should be taken {\it cum grano salis}. However, a variation of the
measured intensity of the 203-GHz line up to 50\% will not introduce
significant changes in our fit results.

Assuming an ortho/para ratio of 3, and a [$^{16}$O]/[$^{18}$O] ratio
of 261$\pm$20 (Whiteoak \& Gardner \cite{whiteoak1981}), we estimate
the \WAT\ abundance to be $\sim 2.3\times10^{-6}$ in this region. This
result yields a [HDO]/[\WAT] ratio of 6.5$\times10^{-4}$, which is in
order-of-magnitude agreement with the estimate of 1.8$\times10^{-4}$
by Gensheimer et al.  (\cite{gensh1996}). The discrepancy between the
[HDO]/[\WAT] ratio calculated by us, and the [D]/[H] ratio measured
towards the Galactic Center (cf. Lubowich et al.
\cite{lubowich2000}), will be discussed in \S~\ref{sec:discussion}.

Knowing the [HDO]/[\WAT] ratio, the o-\WATEIGH\ abundance in Phase II
can be determined\footnote{We can reasonably assume that the gas-phase
  water abundance at this stage be determined mostly by evaporation
  from dust grains or freeze-out onto dust grains, and that these
  processes affect both \WAT\ and HDO equally. In other words, we
  assume the [HDO]/[\WAT] ratio to be constant throughout Phase I and
  II. However, even considering that fractionation is likely to take
  place in the cold portions of the gas, the [HDO]/[\WAT] ratio in the
  Galactic Center is expected to assume a maximum value of $\sim
  10^{-3}$ (E. Bergin, priv.  comm.; also cf. Roberts \& Millar
  \cite{robmill2000}), not very far from our estimate and therefore
  not affecting our conclusions. To make our point even stronger,
  fractionation will be most active in the coldest, outermost regions
  of the envelope, which contribute less than 1\% to the total gas
  column density.}, [o-\WATEIGH]~$\sim 2.1 \times 10^{-10}$. The
resulting total \emph{peak} column density of \WAT\ (hot core + warm
envelope) is $\sim 3.6\times 10^{19}$~\percc.
The 548-GHz o-\WATEIGH\ absorption expected to be produced in the warm
envelope is then predicted, for each core, with our radiative transfer
code. The two spectra are opportunely combined to reproduce the
attenuation of the SWAS beam, and finally compared to the N00 data.

A quantitative comparison between the measured intensities of the
whole set of HDO and \WATEIGH\ lines and the model results can be
found in Tab.~\ref{tab:model_data}. Most of the spectral line data are
reproduced within errors of $\sim 30$\%, with three exceptions
(indicated with a star in Tab.~\ref{tab:model_data}):
\begin{itemize}
\item[1)] The 226-GHz HDO emission line observed towards SgrB2(M) is a
  factor of 2 weaker, and
\item[2)] the 143-GHz HDO line in SgrB2(N) is 4 times stronger than
  predicted by the model. Such discrepancies are likely to be due to
  the intrinsic chemical and physical differences between the
  molecular gas components of the two hot cores (cf. Miao \& Snyder
  \cite{miao1997} and references therein), which have not been taken
  into account in our model.
\item[3)] Most interestingly, the predicted ground-state o-\WATEIGH\ 
  absorption at 548 GHz is much shallower than observed
  (Fig.~\ref{fig:hdo_water}; N00): \emph{the abundance of
    non-deuterated water in the warm envelope is not
    sufficient\footnote{In fact, the column density of o-\WATEIGH\ is
      of the same order of magnitude as that of HDO, but the Einstein
      B coefficient of the o-\WATEIGH($1_{1,0}-1_{0,1}$) transition is
      about a factor of 2 smaller than that of the
      HDO($1_{1,1}-0_{0,0}$) transition, resulting in a shallower
      absorption line.} to produce the absorption feature observed at
    548 GHz.}
\end{itemize}

\begin{table}[htbp]
\caption{Comparison between the measured intensities and opacities of the HDO 
and \WATEIGH\ lines in our sample (Tab.~\ref{tab:data_list}), and the predictions
from the modified Z95b model (\S~\ref{where}). At this stage, only the hot cores and the warm 
envelope of SgrB2 are included in the model. Note that the predicted optical 
depth of the 548-GHz o-\WATEIGH\ line is a factor of 8 smaller than observed 
(see \S~\ref{hotcore} and \S~\ref{hotlayer}).}\label{tab:model_data}
  \begin{center}
    \begin{tabular}{cccl}
\hline
\hline
Line    & Source  & Observed $T_{\rm mb}$ & Model $T_{\rm mb}$   \\
(emission)&         &     (K)               &     (K)              \\
\hline                                                          
HDO 241 &   M     &        --             &      0.9            \\
        &   N     &        1.2            &      0.9             \\
HDO 226 &   M     &        0.4            &      0.8$^{\star}$  \\
        &   N     &        1.3            &      0.8             \\
HDO 143 &   M     &        --             &      0.1             \\
        &   N     &        0.4            &      0.1$^{\star}$    \\
p-\WATEIGH\ 203& N&        1.5            &      1.5             \\
\hline                                                          
(absorption) &         &   Observed $\tau$     &    Model $\tau$      \\
\hline                                                          
HDO 894 &   M     &   $0.4$               & $0.5$              \\
        &   N     &   $0.8$               & $0.6$              \\
o-\WATEIGH\ 548& M+N& 0.8                 & 0.1$^{\star}$      \\
\hline
    \end{tabular}
\begin{list}{}{}
\item[$^{\star}$] deviates by more than 50\% from observed value
\end{list}
  \end{center}
\end{table}

\subsection{A ``hot'' issue: \WAT\ in the hot layer}\label{hotlayer}
The result that the observed \WATEIGH\ absorption cannot, according to
our model, be produced in the warm envelope alone, is a very important
one: first of all, our predictions confirm, at least from a
qualitative point of view, the hypothesis of C02 that the most
important contribution to the observed \WATEIGH\ absorption comes from
the foreground hot gas layer. Secondly, having predicted the
contribution of the warm envelope to the \WATEIGH\ absorption, it is
possible to estimate the spatial width and \WAT\ abundance necessary
for the hot layer to produce the observed feature, given the
observational constraints mentioned in section \S~\ref{where}. In what
follows, we will assume the hot layer to be located right outside of
the warm envelope (see Fig.~\ref{source_model}), although its exact
location does not matter for our purposes.

Because there is no reason to believe it has spherical symmetry, the
hot layer cannot be modeled directly with our radiative transfer code.
However, we can use the radiative transport equation to calculate the
total intensity emerging from the hot layer, $I_{\rm TOT}$, given a
background emission $I_{\rm HC+WE}$ as calculated by our model for the
complex made up of hot core and warm envelope (results listed in
Tab.~\ref{tab:model_data}):
\begin{equation}
I_{\rm TOT} = I_{\rm HC+WE} \cdot e^{- \tau_{\rm HL}} + I_{\rm HL} \cdot (1-e^{-\tau_{\rm HL}}),
\end{equation}
where $\tau_{\rm HL}$ is the optical depth of the hot layer, and
$I_{\rm HL} \cdot (1-e^{-\tau_{\rm HL}})$ is the hot layer emission
corrected for self-absorption. We assume a water abundance, relative
to H$_2$, of $8\times10^{-4}$, based on the abundance of atomic oxygen
not locked in CO
(cf.  Meyer, Jura \& Cardelli \cite{meyer1998} for a determination of
the interstellar O abundance; Shaver et al. \cite{shaver1983} and
Rolleston et al. \cite{rolleston2000} for the parametrization of the
abundance gradient of carbon and oxygen in the Galaxy). The spatial
width of the hot layer, $\Delta w_{\rm HL}$, is treated as a free
parameter, with 0.01~pc~$\leq \Delta w_{\rm HL} \leq 3$~pc, and we
assume the gas sheet to be extended perpendicularly to the line of
sight (see Fig.~\ref{source_model}).

Both dust and gas contribute to $\tau_{\rm HL}$ and $I_{\rm HL}$.
However, it is reasonable to assume the contribution of such a thin,
diffuse layer of dust to be negligible with respect to the background
envelope, and only the emission and opacity of the molecular gas will
be taken into account. The fractional populations of levels $1_{1,0}$
and $1_{0,1}$ of o-\WATEIGH\ have been calculated, with our radiative
transfer code, for a thin sheet of hot gas of density $n({\rm
  H_2})_{\rm HL} = 10^3$~\percc\ and temperature $T_{\rm gas} =
500$~K. Both values match the lower limits derived by C02 through the
modeling of the ammonia absorption features observed with ISO at
infrared wavelengths. A strict upper limit on the H$_2$ density,
$n({\rm H_2})_{\rm HL} \la 5\times10^{3}$~\percc, has been determined
by H\"uttemeister et al.  (\cite{huette1995}) through cm-wavelength
observations of NH$_3$, SiO and HC$_3$N. These values are further
supported by measurements of the intensity of the 691-GHz CO(6-5)
line, performed with the CSO by P.  Schilke \& D.C. Lis (unpublished
data). Also, note that the ratio of the fractional populations of the
o-\WATEIGH\ levels is insensitive to temperature changes in the
500-700~K window indicated by C02.

The best fit of the o-\WATEIGH\ column density is determined based on
the distribution of the $\chi^2$ values given by:
\begin{equation}
\chi^2 = \frac{(\tau_{\rm obs} - \tau_{\rm mod})^2}{\sigma(\tau_{\rm obs})^2},
\end{equation}
where $\tau_{\rm obs}$ and $\tau_{\rm mod}$ are, respectively, the
opacity of the observed 548-GHz \WATEIGH\ absorption, and the
predicted opacity of the feature for each pair of [\WATEIGH]$_{\rm
  HL}$ and $\Delta w_{\rm HL}$ values. $\sigma(\tau_{\rm obs})$ is the
error on the observed line opacity, calculated by adopting a 20\%
error on the continuum level measured by SWAS at 548 GHz (N00). The
minimum $\chi^2$ values yield a o-\WATEIGH\ peak column density of
$\sim 10^{14} \, \cmsq$, hence a \WAT\ column density of $\sim 3.5
\times 10^{16}$~\cmsq, corresponding to $\Delta w_{\rm HL} \sim
0.02$~pc. We consider the value of $\Delta w_{\rm HL}$ to be a lower
limit, but expect it to be correct within a factor of a few ($\sim
2$-3). If we assume a [HDO]/[\WAT] ratio equal to the [D]/[H] ratio
measured towards the Galactic Center ($1.7\times10^{-6}$, Lubowich et
al.  \cite{lubowich2000}) as expected for gas at this temperature,
then the HDO peak column density in the hot layer, $N$(HDO)$_{\rm HL}
\sim 5.2\times10^{10}$~\cmsq, will contribute with 0.3\% only to the
absorption produced by the HDO component located in the warm envelope,
in agreement with our hypothesis (\S~\ref{where}).

An overview on the radial density and temperature profiles adopted to
model the three regions of the cloud complex is given in
Tab.~\ref{model_parameters}. Each profile, $P(r)$, is described by the
equation
\begin{equation}
P(r) = a_0+a_1\cdot\left(\frac{r}{r_0}\right)^{\alpha}.
\end{equation}
Also, Tab.~\ref{tab:summary} summarizes the peak column densities of
H$_2$, HDO and \WAT\ resulting from our model.  The \WAT\ abundance and
the [HDO]/[\WAT] ratio are also stated.

\begin{table*}
\caption{Description of the radial profiles of density and temperature
used to model the hot cores and warm envelope (HC and WE respectively, \S~\ref{hotcore}), 
and the hot layer (HL, see \S~\ref{hotlayer}) 
of the SgrB2 cloud complex. The profiles are described by the coefficients 
$a_0$, $a_1$, $r_0$ and $\alpha$ according to the law: 
$P(r) = a_0+a_1\cdot(r/r_0)^{\alpha}$. The radii are (cf. Fig.~\ref{source_model}): $r < 0.05$~pc for the hot core and 
$0.05 \, {\rm pc} \leq r \leq 22.5$~pc for the warm envelope. The hot layer is, in our model, 
0.02~pc thick. }\label{model_parameters}
  \begin{center}
    \begin{tabular}{cccccccccccccccc}
\hline
\hline
 & & \multicolumn{4}{c}{$n({\rm H_2})$} & & \multicolumn{4}{c}{$T_{\rm gas}$} & & \multicolumn{4}{c}{$T_{\rm dust}$}  \\
 & & $a_0$    & $a_1$ & $r_0$ & $\alpha$ &  &$a_0$    & $a_1$ & $r_0$ & $\alpha$ & &$a_0$    & $a_1$ & $r_0$ & $\alpha$\\
Region& &(\percc)&(\percc)&(pc)& & &(K)&(K)&(pc)& & &(K)&(K)&(pc)& \\
\cline{1-1} \cline{3-6} \cline{8-11} \cline{13-16}
HC & & $3.4\times10^7$ & 0 & -- & -- & & 200 & 0 & -- & -- & & 200 & 0 & -- & -- \\
WE & & $2.2\times10^3$ & $5.5\times10^4$ & $1.25$ & $-2$ & & 0 & 40 & 1 & $-0.5$ & & 0 & 40 & 1 & $-0.5$ \\
HL & & $10^3$ & 0 & -- & -- & & 500 & 0 & -- & -- & & -- & -- & -- & -- \\
\hline
    \end{tabular}
  \end{center}
\end{table*}

\begin{table*}[htbp]
\caption{Summary of the peak column densities of H$_2$, HDO and \WAT\ derived 
in \S~\ref{hotcore} and \ref{hotlayer}. The values relative to 
hot cores and warm envelope (HC+WE) are based on the radiative transfer code illustrated in Z95b and 
modified as in \S~\ref{molnd}. The determination of the hot layer values is 
described in \S~\ref{hotlayer}. The [HDO]/[\WAT] ratio for each region is also indicated.
The value of [HDO]/[\WAT] in the hot layer is assumed equal to the [D]/[H] ratio in the Galactic 
Center (cf. Lubowich et al. \cite{lubowich2000}).}\label{tab:summary}
  \begin{center}
    \begin{tabular}{ccccc}
\hline
\hline
              & $N({\rm H_2})$   &  $N({\rm HDO})$  & $N({\rm \WAT})$   & [HDO]/[\WAT]  \\
Region        &   (\cmsq)        &   (\cmsq)        &  (\cmsq)          &     \\
\hline
HC+WE &$1.9\times 10^{25}$ &$2.3\times10^{16}$&$3.6\times10^{19}$&$6.4\times10^{-4}$       \\     
HL &$6\times10^{19}$&$5.2\times10^{10}$&$3.5\times10^{16}$&$1.7\times10^{-6}$$^{\mathrm{(\star)}}$\\                 
\hline
\end{tabular}
\begin{list}{}{}
\item[$^{\mathrm{(\star)}}$] assumed (cf. Lubowich et al. \cite{lubowich2000}) 
\end{list}
\end{center}
\end{table*}

\section{Discussion}\label{sec:discussion}
The unambiguous identification of the hot layer as the gas component
responsible for the 65-\kms\ water absorption towards SgrB2 is
definitely the main result of this work. This confirms the general
picture proposed by C02. However, our model provides, for the hot
layer, a value of the water column density, $N(\WAT)_{\rm HL}$, which
is a factor of $\sim 10$ lower than that inferred by C02, due to the
fact that we assumed a lower H$_2$ density.  In fact, previous
estimates of the column density of the absorbing water (cf.  Z95a,
N00, and \S~\ref{subsect:hdo_water_ratio} of this paper) are in very
good agreement with our result of $N(\WAT)_{\rm HL} \sim 3.5 \times
10^{16}$~\cmsq: the assumption employed in these estimates that only
the ground state of \WAT\ be populated \emph{is correct} for the hot
layer, although it does not apply to the warm envelope. However, the
column density of water actually responsible for the observed
absorption is negligible, with respect to the total column density
estimated over the whole envelope (see Tab.~\ref{tab:summary}). It is
important to keep in mind that column densities calculations from
ground-state absorption features, if based on the ''classical''
assumptions mentioned in \S~\ref{subsect:coldens}, can be highly
misleading in regions as complex as SgrB2, and are likely to yield
underestimates of several orders of magnitude (a factor of $\sim 1000$
in our case).

As previously mentioned, several heating mechanism have been proposed
to explain the existence of such a hot, relatively thin sheet of
molecular gas. The origin of the whole cloud complex has been proposed
to be linked to large-scale cloud-cloud collisions (Hasegawa et al.
\cite{hasegawa1994}), so shock-induced heating is definitely possible,
although it remains to be explained why the velocity of the hot layer
is identical to that of the warm envelope.  Mart\'{\i}n-Pintado et al.
(\cite{mpintado2000}) argue rather in favour of an X-ray-driven
chemistry, based on the observed spatial correlation of the Fe 6.4-keV
emission line with the SiO(1-0) emission at 43.4 GHz. In a recent
paper by Goicoechea \& Cernicharo (\cite{goicoechea2002}), a
[OH]/[\WAT] abundance ratio of 0.1-1 is estimated for the hot layer,
which, according to the authors, points towards the presence of a
strong UV field illuminating the outer shells of the cloud.  The high
abundance of \WAT\ estimated by us does not allow to discriminate
among the proposed mechanisms, and it is anyway likely that all of
them contribute, to some extent, to the anomalous heating of this
region. However, our radiative transfer calculations allow us to
separate the chemistry driving the water abundance in the hot layer
from that taking place in the warm envelope, thus helping to set more
solid constraints on the physical characteristics of the hot diffuse
gas, such as, for example, its spatial extent.

We would also like to stress the high deuterium fractionation
([HDO]/[\WAT]~$\sim 6.4 \times 10^{-4}$) inferred, for the SgrB2 hot
cores, by our radiative transfer model. Our value is in agreement
(about a factor of 3 higher) with that estimated by Gensheimer et al.
(\cite{gensh1996}) for SgrB2(N), and it is almost 400 times larger
than the elemental [D]/[H] ratio measured towards the Galactic
Center\footnote{The ratio of the abundance of a deuterated species
  with respect to its non-deuterated counterpart, [XD]/[XH], is
  expected to scale with the [D]/[H] ratio in the gas (Gerin \& Roueff
  \cite{gerin1999}; E. Bergin, priv. comm.).}  (Lubowich et al.
\cite{lubowich2000}). An enhancement of the deuterium fractionation in
hot-core-type sources has been observed for a variety of chemical
species ([\METHD]/[\METH], Mauersberger et al.  \cite{mauers1988};
[\DAMM]/[\AMM], Jacq et al.  \cite{jacq1990}; [DCN]/[HCN], Schilke et
al. \cite{schilke1992}, Hatchell, Millar \& Rodgers
\cite{hatchell1998}; [\DCOP]/[\HCOP], Jacq et al.  \cite{jacq1999};
[HDS]/[\HTWS], Hatchell, Roberts \& Millar \cite{hatchell1999};
[CH$_2$DCN]/[CH$_3$CN], Gerin et al. \cite{gerin1992}; and
[HDO]/[\WAT], Jacq et al.  \cite{jacq1990}, \cite{jacq1999};
Gensheimer et al. \cite{gensh1996}; Pardo et al. \cite{pardo2001}),
and is generally attributed to evaporation of deuterated species from
grain mantles due to the formation of an embedded heating source (see
Walmsley et al.  \cite{walmsley1987}). It must be kept in mind that
the quoted estimates of the [D]/[H] ratio in the Galactic Center are
affected by large uncertainties (about 1 order of magnitude). However,
we can assume our result to be in agreement with the general finding
that the abundance ratio of deuterated species to their non-deuterated
counterparts is, in hot-core-type sources (cf.  references above),
enhanced by a factor of a few $10^2$ with respect to the elemental
[D]/[H] ratio. Thus, the value of [HDO]/[\WAT] derived in
\S~\ref{hotcore} yields a [D]/[H] ratio of a few $10^{-6}$, supporting
the hypothesis of a lower [D]/[H] ratio in The Galactic Center region.

\begin{acknowledgements}
  
  The authors are grateful to D. Neufeld for providing the 548-GHz
  SWAS data in digital format; to M. Walmsley, T. Wilson and C.
  Ceccarelli for their valuable comments; and to the referee, E.
  Bergin, whose input has very much contributed to improve the quality
  of this paper. CC acknowledges travel support and sunny hospitality
  from the Submillimeter Wave Astrophysics Research Group at Caltech.

\end{acknowledgements}

\end{document}